\documentclass{ws-procs10x7}
\usepackage{balance,epsfig}

\newcolumntype{d}[1]{D{.}{.}{#1}}

\def\draftnote{}

\makeindex
\begin{document}

\title{Recent Developments in Soft-Collinear Effective Theory}

\author{Th.~Feldmann} 

\address{Fachbereich Physik,
   Universit\"at Siegen, D-57068 Siegen, Germany}


\twocolumn[\maketitle\abstract{
Soft-collinear effective theory provides a systematic
theoretical framework to describe the factorization of
short- and long-distance QCD dynamics in hard-scattering
processes that contain both, soft and energetic particles/jets. 
I present a short guide to recent theoretical achievements 
and to phenomenological applications in heavy $B$-meson decays.}
\keywords{QCD; Factorization; $B$\/-meson decays.\hfill
{\tt Siegen Preprint SI-HEP-2006-11.}}
]


\begin{figure*}[pbt]
\centerline{
(a) \hspace{-1.7em} \psfig{file=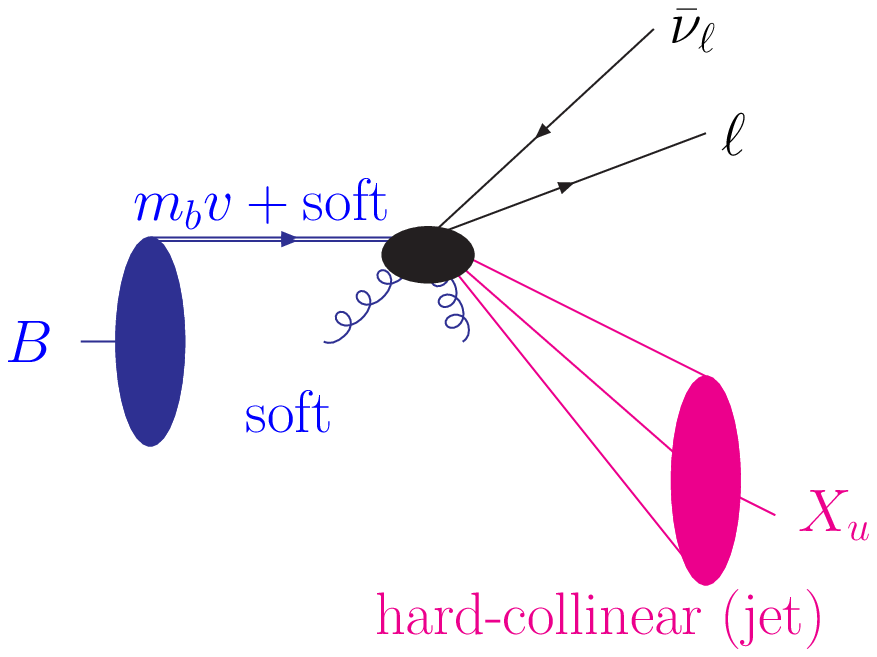,width=0.3\textwidth} \ \ \
(b) \hspace{-1.7em} \psfig{file=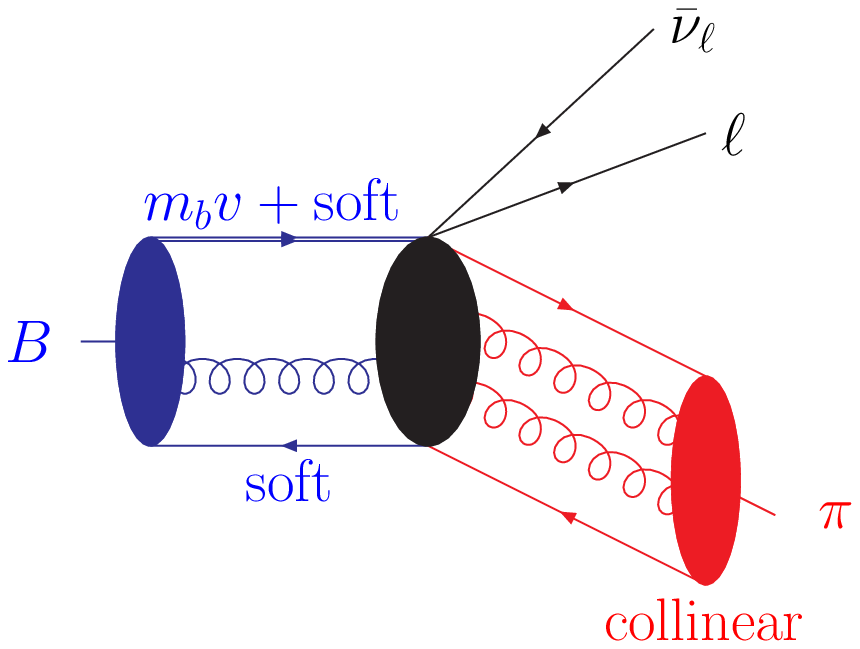,width=0.3\textwidth} \ \ \
(c) \hspace{-1.7em} \psfig{file=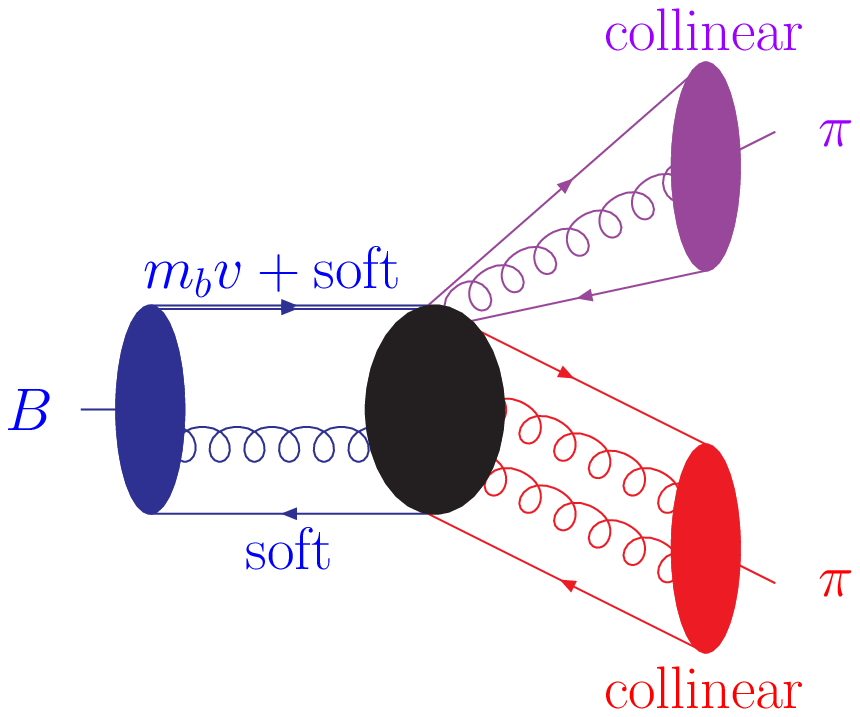,width=0.3\textwidth} 
}
\caption{Examples for momentum configurations in $B$\/-decays: (a) inclusive decay $B \to X_u \ell \nu$
 in the shape function region,
 (b) semi-leptonic decay
 $B \to \pi \ell \nu$ at large recoil, (c) 
 charm-less non-leptonic decay $B\to \pi \pi$.}
\label{fig:regions}
\end{figure*}

\section{Introduction}

The idea of factorization in high-energy processes
is to isolate (perturbative) short-distance QCD dynamics 
from long-distance (non-perturbative) hadronic effects.
A well-known example is the weak effective Hamiltonian
which describes quark and lepton flavour transitions 
in terms of local operators $O_i(x)$, 
multiplied by Wilson coefficients $C_i(\mu)$,
containing the short-distance dynamics above a scale
$\mu \leq M_W$. In $B$\/-meson decays 
the theoretical task remains to calculate hadronic 
matrix elements
$
  \langle X| O_i(x) |B \rangle_{\mu \sim m_b} \,,
$
with $|X\rangle$ an exclusive or inclusive final state.
As the mass of the decaying  
$b$\/-quark is large compared to the QCD scale,
$m_b \gg \Lambda$, QCD dynamics which involves
virtualities of the order $m_b^2$ (``hard modes'') can
still be treated perturbatively. 

In the following, we will concentrate on $B$\/-meson
decays where the final state consist of hadronic systems 
with large energy but small invariant mass. 
Three typical examples are shown in 
Fig.~\ref{fig:regions}. In these cases a second perturbative
scale $\mu_{\rm hc} \sim \sqrt{\Lambda m_b}\gg \Lambda$ appears.
In inclusive decays, it is related to the invariant
mass of the hadronic jet. In exclusive decays, these (``hard-collinear'')
modes arise from the interaction of soft spectators in the
$B$\/-meson and energetic (``collinear'') degrees of freedom in
the final state. 
The remaining non-perturbative
dynamics is encoded in hadronic matrix elements of {\em non-local}\/
operators. For inclusive $B$\/-decays, these define so-called
shape functions (SFs) which describe the momentum distribution of the
$b$\/-quark in the $B$\/-meson due to soft gluon interactions. 
In exclusive decays, light-cone distribution
amplitudes (DAs) for the $B$\/-meson and its decay products appear.

In the effective-theory approach, 
the factorization theorems can be obtained from
a 2-step matching procedure, based on a simultaneous expansion
in $\alpha_s$ and $\Lambda/m_b$.\cite{Bauer:2000yr,Beneke:2002ph,Chay:2002vy,Hill:2002vw}
In the first step, one integrates out hard modes from
QCD to derive a soft-collinear effective theory (SCET$_{\rm I}$) where
soft and hard-collinear degrees of freedom interact. The renormalization
group for operators in SCET$_{\rm I}$ is used to evolve 
the coefficient functions down to the hard-collinear scale,
resumming logarithms $\ln [m_b/\mu_{\rm hc}]$.
Then, in inclusive
decays, one integrates out the hard-collinear (jet) modes applying
quark-hadron duality, ending up with the usual heavy-quark effective
theory (HQET). In exclusive decays, one is left with collinear 
degrees of freedom, whose interactions with
soft modes is described by a so-called SCET$_{\rm II}$.

\section{Inclusive Decays (SF region)}

Experimental studies of inclusive $B$ decays often require certain
kinematic cuts which imply sensitivity to the shape-function
region as described above\cite{Neubert:1993ch,Bigi:1993ex}.
A prominent example is $B \to X_u\ell \nu$, where the
spectrum in the variable $P_+ = E_X - |\vec P_X|$ has
to be restricted to values of $P_+ < \Delta \leq M_D^2/M_B$
in order to suppress the charm background.
The effective-theory treatment of the partial rate
leads to the factorization theorem,\cite{Bauer:2003pi,Bosch:2004th}
\begin{eqnarray}
&& \frac{d\Gamma_u}{dP_+} 
\propto
  \int\limits_0^1  dy \, y^{-a} \, H_u(y,m_b) \, U(m_b,\mu_{\rm hc}) \,
\nonumber \\[0.0em]
&& {} \times   \int\limits_0^{P_+} d\hat \omega \,   
 J(y (P_+ - \hat
  \omega),\mu_{\rm hc} ) \, \widehat S(\hat \omega,\mu_{\rm hc})\,,
\end{eqnarray}
valid to leading power in  $\Lambda/m_b$.
Here the hard function $H_u(y)$ can be calculated perturbatively
in QCD ($y=(E_X+|P_X|)/m_b$). 
The jet function $J(u)$ is calculated in SCET$_{\rm I}$,
and $\widehat S(\hat \omega)$ is the leading shape function
in HQET.
The renormalization group functions $U(m_b,\mu_{\rm hc})$ and
$a=a(m_b,\mu_{\rm hc})$ resum logarithms between the hard and
hard-collinear factorization scale.
An analogous factorization theorem holds for photon-energy spectra
in $B \to X_s \gamma$.

Shape-function effects also have to be considered in inclusive
$B \to X_s \ell^+ \ell^-$ decays in the region of small invariant
lepton-pair mass. Here the experimental studies require an
additional cut on the hadronic invariant mass to eliminate
combinatorial background like $b \to c\ell\nu \to s\ell\ell\nu\nu$.
The description within SCET has recently been discussed in
Ref.~\cite{Lee:2005pk}.

\subsection{Theoretical status}

Besides the hard matching coefficients 
(known to NLO for $b \to u\ell \nu$ and $b \to s\gamma$),
there has been recent progress in the evaluation of the jet function
which is now known to NNLO accuracy\cite{Becher:2006qw} for massless
quarks. 
The two-loop evolution kernel for the leading-power 
$B$\/-meson shape function has been derived in 
Ref.\cite{Becher:2005pd}. Sub-leading shape
functions, which enter at order $\Lambda/m_b$ in SCET, 
have been classified in 
Refs.\cite{Lee:2004ja,Bosch:2004cb,Beneke:2004in,Tackmann:2005ub} 
for the massless case, and in Ref.\cite{Boos:2005by}
for the massive case.

\subsection{SF-independent relations}

From the factorization formulas
for $B \to X_s\gamma$ and $B \to X_u\ell\nu$ 
one can derive a SF-independent relation
between partially integrated spectra,\cite{Lange:2005qn} (see also 
Refs.\cite{Leibovich:1999xf,Hoang:2005pj,Paz:2006me})
\begin{eqnarray}
&&  \Gamma_u(\Delta) \equiv 
\int\limits_0^\Delta d P_+ \, {\frac{d\Gamma_u}{dP_+}} =
\cr
&& |V_{ub}|^2 \, \int\limits_0^\Delta \!\! dP_+ \, {W(\Delta,P_+)} \, 
 {\frac{1}{\Gamma_s} \, \frac{d\Gamma_s}{dP_+}} \,,
\end{eqnarray}
where, to leading power in $\Lambda/m_b$,
the weight function $W(\Delta,P_+)$ is 
perturbatively calculable.
Choosing $\Delta \sim 650$~MeV, and considering the two
experimental spectra, one can determine the CKM element
$|V_{ub}|$ in the SM with rather good accuracy. The analysis
in Ref.\cite{Lange:2005qn} based on the two-loop result
for $W(\Delta,P_+)$ predicts $\Gamma_u(0.65~{\rm GeV}) =
 (46.5 \pm 4.1) \, |V_{ub}|^2 \, {\rm ps}^{-1}$, where the
quoted number also includes an estimate of power corrections.
Situations with  more complicated cuts
have been studied in Ref.\cite{Lange:2005xz}.

Similar relations can be derived from the $B \to X_c\ell \nu$
decay\cite{Mannel:2004as}, if one treats the charm-quark as 
a massive quark in SCET$_{\rm I}$ (assuming 
the power counting $m_c^2 \sim m_b \Lambda$).
Here one considers the spectral variable 
$U= P_+ - m_c^2/y m_b^2$ in the region of phase space where
$U \sim \Lambda$. The comparison of 
$d\Gamma_c/dU$ and $d\Gamma_u/dP_+$ gives an estimate of
$|V_{ub}/V_{cb}|$. The one-loop weight function for this case
has been calculated in Ref.\cite{Boos:2005qx}. Power corrections are
potentially large. The experimental analysis
of the $U$\/-spectrum in $B \to X_c\ell \nu$ is still to be
performed.

\section{Exclusive Decays (large recoil)}

Factorization theorems also exist for
exclusive $B$\/-meson decays with large energy transfer to
the final-state hadrons. For instance, 
the amplitudes for charmless non-leptonic two-body $B$ decays
can be written as\cite{Beneke:1999br}
\begin{eqnarray}
&& {\cal A}_i(B \to M M') =
 {\xi_{M}} \cdot
  C_i^{\rm I} \otimes {\phi_{M'}} \, 
\nonumber \\[0.1em] &&
\qquad {}
 + T_i^{\rm II}  \otimes {\phi_B} \otimes {\phi_M}
  \otimes {\phi_{M'}} \,,
\label{eq:qcdf}
\end{eqnarray}
up to power corrections of order $\Lambda/m_b$.
Here  the distribution amplitudes
$\phi_{B,M,M'}$ parameterize the non-perturbative dynamics of
the two-quark Fock state in the respective hadron.
The short-distance coefficient
$C_i^{\rm I}$ stems from
the perturbative matching between QCD and SCET$_{\rm I}$.
The spectator term $T_i^{\rm II}$ further factorizes into a hard
function $C_i^{\rm II}$ and a jet function in SCET$_{\rm I}$.
A peculiarity of exclusive $B$ decays is the appearance of ``non-factorizable''
input functions, in the above case the universal form factor
$\xi_M$ for $B \to M$ transitions. 
In these
objects the soft and collinear degrees of freedom in SCET$_{\rm II}$
cannot be completely separated (at least not with standard perturbative
methods). 
Leading-power factorization proofs exist for the somewhat simpler
cases of $B \to \gamma\ell\nu$\cite{Lunghi:2002ju,Bosch:2003fc},
$B \to \pi(\rho)\ell\nu$\cite{Beneke:2003pa,Lange:2003pk}, and
$B\to K^*(\rho)\gamma$\cite{Becher:2005fg}. 
Applications of QCD factorization
for $B \to K^*(\rho) \ell^+\ell^-$ decay observables can be
found in Refs.\cite{Beneke:2001at,Grinstein:2005ud,Ali:2006ew}.
For the status of perturbative 
calculations of the short-distance coefficient 
functions\cite{Beneke:2004rc,Hill:2004if,Kirilin:2005xz,Beneke:2005gs}
see Ref.\cite{Jaeger:2006} in these proceedings.

Factorization theorems have also been 
formulated\cite{Cirigliano:2005ms}
for the exclusive semi-leptonic radiative decay $B \to \pi\gamma\ell\nu$.
In the phase-space region where the pion is soft and the
photon is energetic, the leading contribution to the
decay amplitude factorizes as
${\cal A} = H \cdot J \otimes S(B \to \pi)$. Here, the
soft function $S$ is the generalized parton
distribution for $B \to \pi$ transitions\cite{Feldmann:1999sm}.
Factorization-based numerical predictions
for photon spectra and angular distributions differ from
popular Monte Carlo models for real-photon corrections 
to $B\to \pi \ell \nu$.

\begin{table*}[t!!!]
\tbl{Comparison of different phenomenological assumptions in
BBNS and BPRS approaches.
\label{tab:comp}}
{\begin{tabular}{p{0.25\textwidth}| p{0.33\textwidth} p{0.33\textwidth}}
\toprule
 & {BBNS} & {BPRS} \\
\colrule
{charm penguins}
  & included in hard functions 
  & complex fit parameter $\Delta_P$
\\
{spectator term} & perturbative factorization    
& fit to data 
\\
{ext.\ hadronic input}
   & form factor and LCDAs \newline 
    {\footnotesize (different scenarios)} 
   & LCDA for light meson, only
\\
{power corrections}
  &  model-dependent estimate \newline
    {\footnotesize (complex functions $X_A$ and $X_H$)}
  &  $\to$ systematic uncertainties \newline {\footnotesize (unspecified)}
\\
\botrule
\end{tabular}}
\end{table*}

\subsection{BBNS vs.\ BPRS approach}

The prescription of non-factorizable power corrections to 
(\ref{eq:qcdf}) requires additional hadron\-ic parameters which,
at present, cannot be estimated in a systematic way. 
An important example are strong phases from 
final-state rescattering. 
The factorization formula predicts these phases
to be either perturbative (and calculable) or (formally)
power-suppressed.
Different assumptions about non-factorizable effects 
thus lead to different theoretical predictions for
exclusive $B$\/-decays. 
Two popular examples are the ``BBNS
approach''\cite{Beneke:2001ev} and
the ``BPRS approach''\cite{Bauer:2004tj}.
A qualitative comparison is given in Table~\ref{tab:comp}. 
A (controversial) discussion 
can be found in Refs.\cite{Beneke:2004bn,Bauer:2005wb}.

\subsection{Enhanced electroweak penguins in $B \to VV$}

An advantage of the effective-theory framework is the
definite power counting assigned to fields and operators
appearing in the effective Lagrangian. An interesting
application for $B$ decays
to two light vector mesons has recently been pointed out 
in Ref.\cite{Beneke:2005we}
(see also Ref.\cite{Lu:2006nz} for more examples). 
From the $(V-A)$ structure of weak
interactions one would naively expect the helicity amplitudes to
scale as $A_0 : A_- : A_+ = 1 : \Lambda/m_b : \Lambda^2/m_b^2$. 
The inclusion of the electromagnetic penguin operator ${\cal O}_7^\gamma$ 
via QED corrections to $T_i^{\rm I}$ is shown to {\em enhance}\/ the
transverse helicity amplitudes by a factor $m_b^2/m_V^2$ which 
can compensate for the electromagnetic suppression factor $\alpha_{\rm em}$. 
Among others, this implies a higher sensitivity 
to $(V+A)$ structures in certain new-physics models than naively
anticipated.

\subsection{SCET sum rules}

The non-factorizable form factor $\xi_M$ can be
estimated from sum rules in the effective theory
SCET$_{\rm I}$. Here, one considers a
correlation function where the energetic meson in the
final state is replaced by an appropriate current\cite{DeFazio:2005dx}
(see also Ref.\cite{Khodjamirian:2005ea}). 
The correlation function in the Euclidean region does
factorize into a hard-collinear short-distance kernel
(known to NLO\cite{DeFazio:2005dx})
and a soft $B$\/-meson DA.
\begin{eqnarray}
  \Pi(P_+) &=& \int\limits_0^\infty d\omega \, 
               T(\omega,P_+) \, \phi_-^B(\omega) + \ldots
\end{eqnarray}
Inserting the result into a dispersion
relation, one obtains a sum rule for the form factor $\xi_M$,
which depends on the parameters used to describe the continuum
contribution to the spectral function. The dependence on the
sum-rule parameters and the (at present) not so well-known
$B$\/-meson DA dominate the theoretical uncertainty.

\section{Summary and Outlook}

Factorization of short- and long-distance QCD effects
via effective-theory methods in SCET provides a systematic
and well-defined framework to describe $B$\/-decays into
energetic hadrons. For inclusive decays precise theoretical
predictions can be obtained on the basis of NNLO calculations.
These can be used to determine the CKM element
$|V_{ub}|$ and to constrain new-physics contributions
to $B \to X_s\gamma$ and $B \to X_s \ell^+\ell^-$
(see also Ref.\cite{Hurth} in these proceedings).

Theoretical predictions for exclusive observables are plagued
by non-factorizable contributions which are difficult to
estimate, in particular concerning the strong rescattering
phases in non-leptonic  $B$\/-decays. Sum rules in SCET provide
a promising tool to estimate simple objects, like the soft
$B \to \pi$ form factor $\xi_\pi$. There are also 
attempts to reduce non-factorizable matrix elements to
more fundamental objects in SCET$_{\rm II}$ by introducing
an additional factorization prescription in rapidity 
space.\cite{Manohar:2006nz} How this procedure can
be applied beyond fixed-order perturbation theory is still to 
be worked out.

SCET techniques 
have also been used to describe other high-energy observables, like
jet distributions, or DIS and Drell-Yan near the end 
point.\cite{Lee:2005vs,Bauer:2006qp,Pecjak:2005uh,Manohar:2003vb,Idilbi:2005ky,Chay:2005rz}
Finally, SCET has been combined with non-relativistic QCD to improve the
description of quarkonium production and decay spectra.\cite{Fleming:2003gt,GarciaiTormo:2005ch}

\end{document}